# Modulation of DNA loop lifetimes by the free energy of loop formation

Yi-Ju Chen[a,1], Stephanie Johnson[b,c,1], Peter Mulligan[d,1], Andrew J. Spakowitz[d], and Rob Phillips[e,f,2]

Departments of [a]Physics, [b]Biochemistry and Molecular Biophysics, [e]Applied Physics, and [f]Biology, California Institute of Technology, Pasadena, CA 91125; [c]Department of Biochemistry and Biophysics, University of California, San Francisco, CA 94158; and [d]Department of Chemical Engineering, Stanford University, Stanford, CA 94305

Storage and retrieval of the genetic information in cells is a dynamic process that requires the DNA to undergo dramatic structural rearrangements. DNA looping is a prominent example of such a structural rearrangement that is essential for transcriptional regulation in both prokaryotes and eukaryotes, and the speed of such regulations affects the fitness of individuals. Here, we examine the in vitro looping dynamics of the classic Lac repressor gene-regulatory motif. We show that both loop association and loop dissociation at the DNA-repressor junctions depend on the elastic deformation of the DNA and protein, and that both looping and unlooping rates approximately scale with the looping J factor, which reflects the system's deformation free energy. We explain this observation by transition state theory and model the DNA–protein complex as an effective worm-like chain with twist. We introduce a finite protein–DNA binding interaction length, in competition with the characteristic DNA deformation length scale, as the physical origin of the previously unidentified loop dissociation dynamics observed here, and discuss the robustness of this behavior to perturbations in several polymer parameters.

protein-mediated DNA looping | loop-modulated kinetic rates | transition state

**M**any key cellular processes, from gene regulation to metabolism, require the coordinated physical interaction of biological macromolecules. A classic example is the coordination of DNA and proteins in DNA loop formation, which is a recurring design principle from viruses (1) to animals (2). Despite their prevalence, many questions remain about how these loops form and function in vitro and in vivo (3, 4). Here, we ask how the mechanics of the protein and DNA in a protein-mediated loop govern looping and unlooping dynamics, an issue that has only recently begun to be explored (5).

Recent advances in single-molecule techniques have allowed precise quantification and deeper understanding of the physical properties of DNA (6, 7), and we have taken advantage of one such technique here, in combination with a classic bacterial looping protein. The polymer physics implications of this work, however, are more general than the particular system we focus on. The looping protein we examine here, the Lac repressor, so named because of its role in repressing transcription at the *lac* promoter in the bacterium *Escherichia coli*, was one of the first-described examples of a genetic regulator that acts through specific DNA–protein interactions. Looping is accomplished by means of two DNA binding domains per Lac repressor molecule, allowing it to bind two of its specific DNA sites ("operators") simultaneously, with the intervening DNA adopting a looped conformation (8), as shown schematically in Fig. 1*A*. To measure loop formation and breakdown, we make use of the in vitro single-molecule tethered particle motion (TPM) assay (9–12), a simple but powerful technique for investigating DNA–protein interactions. In TPM, a micrometer-sized bead is tethered to one end of a linear DNA, with the other end attached to a microscope coverslip (Fig. 1*B*). The motion of the bead gives a readout of the effective length of the bead's DNA "tether," such that loop formation induced by the binding of the Lac repressor to its two operators results in a quantifiable reduction of the bead's motion. We record the trajectory of looping and unlooping for each DNA molecule as a function of time in thermal equilibrium (Fig. 1*C*).

These trajectories contain a wealth of information about the DNA–protein interactions in our system. One such quantity that we will focus on here is the looping $J$ factor, $J_{loop} = (1\,\text{M})e^{-\beta \Delta F}$, which encapsulates the thermodynamic cost $\Delta F$ to deform the DNA (and possibly the protein) into the looped conformation, related to the cyclization $J$ factor often used to measure the flexibility of DNA in vitro (13). We previously measured $J_{loop}$ for looping constructs generated by a library of DNAs with different loop lengths and sequences (Fig. 1*C* and *SI Appendix*, Fig. S1), to examine how DNA mechanics affect the energetics of loop formation (12, 14). We showed that this library of constructs allowed us to tune $J_{loop}$ over two orders of magnitude.

Here, we will instead focus on the looping and unlooping "lifetimes" (durations; Fig. 1*C*) for this same library of DNAs. We find that the loop breakdown process at the DNA–protein interface is sensitive to the whole loop's deformation, with both looping and unlooping kinetics exhibiting rather simple forms of scaling with the looping free energy. Such a dependence has not, to our knowledge, been previously reported experimentally (10, 15) or considered in standard physical models for DNA looping (16–20), and suggests DNA looping as a member of a broader class of phenomena where applied force (21–25) or internal stress stored in polymers (5, 7, 26, 27) modulates biochemical reaction rates. Moreover, this result implies possible influence of DNA mechanics on evolution, because both the speeds of turning gene regulation "on" and "off" may be critical for fitness. We provide an explanation for the molecular origins of this dependence and develop a theory of looping kinetics, allowing us

> **Significance**
>
> Storage and retrieval of the genetic information in cells is a dynamic process that requires the DNA to undergo dramatic structural rearrangements. One prominent but still incompletely understood example of such rearrangements is protein-mediated DNA looping. Here, we use a single-molecule biophysical technique to show that the elasticity of the DNA and protein affects the kinetics of both loop formation and, unexpectedly, loop breakdown, and we develop a theory based on polymer physics to explain the origin of these observations. Our results demonstrate a previously unidentified way to probe the looping reaction pathway and quantify the effects of both the DNA and protein elasticity in looping kinetics, and therefore better understand their roles in many gene-regulatory processes.

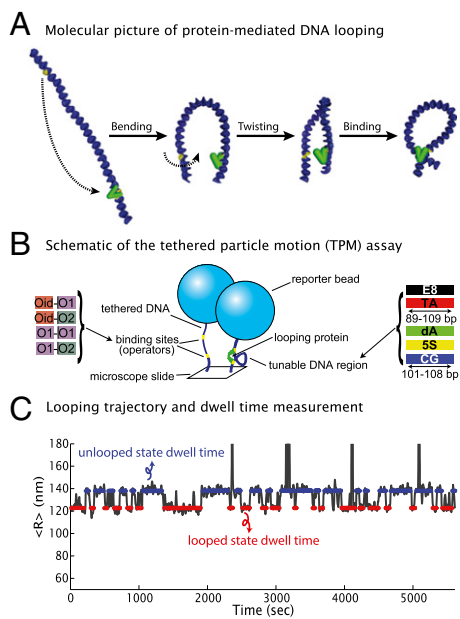

**Fig. 1.** DNA looping dynamics measured by tethered particle motion. (*A*) Loop formation requires the DNA chain to bend and twist to bring the binding sites together and properly orient them. (*B*) The TPM setup, in which single DNA molecules tether microscopic beads to a slide. Looping due to the Lac repressor binding the two operators on the DNA reduces the bead's motion. (*C*) Sample TPM trajectory, ⟨R⟩ versus time, recorded from a single tether and segmented into unlooped (blue) and looped (red) states. The lifetime of a state is how long a trajectory remains in that state before transitioning to a different one.

to probe experimentally inaccessible details of the looping pathway and the looping transition state.

### Results and Discussion

In this work, we use the common single-molecule analysis technique of half-amplitude thresholding (see details in *SI Appendix*, section S1) to obtain distributions of the amount of time each TPM tether spends in the looped or unlooped state, called looped or unlooped lifetimes. We begin by developing a simple kinetic framework for understanding what the measured state lifetimes tell us about the underlying physics of the system, with the basic elements given in Fig. 2. As discussed in more detail in *SI Appendix*, section S2.1.2–4, using standard kinetic analyses we can express the mean unlooped state lifetime, $\langle \tau_{unlooped} \rangle$, in terms of the repressor concentration $[R]$ and the rate constants diagrammed in Fig. 2, as follows:

$$\langle \tau_{unlooped} \rangle = \frac{\left(1 + \frac{k_{off}^A}{[R]k_{on}^A}\right)\left(1 + \frac{k_{off}^B}{[R]k_{on}^B}\right)}{k_{on}^\alpha \frac{k_{off}^A}{[R]k_{on}^A} + k_{on}^\beta \frac{k_{off}^B}{[R]k_{on}^B}}. \quad [1]$$

Note that $\langle \tau_{unlooped} \rangle$ contains two different kinds of rates: $\{k_{on/off}^A, k_{on/off}^B\}$, for the binding/unbinding of the first repressor head to the DNA (we are distinguishing between binding and unbinding to operator $A$ versus operator $B$, because several operators with different affinities for repressor have been described), as well as $\{k_{on}^\alpha, k_{on}^\beta\}$ for the binding of a second operator when the repressor has already bound the first one, which we here allow to differ from the rates for the initial binding event (Fig. 2). On the other hand, $\langle \tau_{looped} \rangle = 1/(k_{off}^\alpha + k_{off}^\beta)$ contains only the loop-affected dissociation rates $\{k_{off}^\alpha, k_{off}^\beta\}$, which we made no a priori assumptions and allow to differ from the simple unbinding events $\{k_{off}^A, k_{off}^B\}$. Experimental measurements of the unlooped and looped lifetimes then tell us how looping affects $k_{on}^{\alpha/\beta}$ and $k_{off}^{\alpha/\beta}$, respectively. Regulation of association rates by flexible linkers and polymer ring closure rates have been discussed in the framework of confined diffusion (16–20), and the effect of confined diffusion from the elastic DNA-repressor loop is likely to dominate $\{k_{on}^\alpha, k_{on}^\beta\}$ in our case as well. However, dissociation rates are usually thought of as local phenomena and dependent only on the interaction strength at the molecular interface, a hypothesis implicitly used in previous works on DNA looping kinetics (5, 10, 15, 28). In contrast, in force spectroscopy experiments, an applied force changes a reaction free energy profile by adding a linear term to it. As a result, the equilibrium constant of the reaction, as well as both the on and off rates (e.g., association and dissociation of chemical bonds, folding and unfolding of RNA or nucleosomes), depend on the pulling force (21–25). With our kinetic measurements, we can address the question of to what degree the dissociation process (i.e., the looped lifetime) is simply a local interaction and to what degree it is affected by the elastic deformation energies of the protein and DNA chain.

As exemplified by Fig. 3A for sequence dA (Fig. 1B), the unlooped and looped state lifetimes extracted from our TPM data show a modulation with loop length, just as the J factor (equivalently, the deformation free energy of the system) does. The other sequences are similarly plotted in *SI Appendix*, Fig. S11, and exhibit more complex behavior when loop length varies more than one helical repeat. Although the unlooped and looped state lifetimes are complicated functions of the loop length and DNA sequence, they are approximately monotonic when plotted versus the J factor, as shown in Fig. 3B. Moreover, this behavior is roughly independent of both loop sequence and, within the range of lengths examined here, loop length, as shown in Fig. 3 *C* and *D*: unlooped and looped state lifetimes for five different sequences spanning one to two helical periods of DNA all follow the same trend with J. Because J is known to be a function of loop length and DNA sequence, it can be viewed as encompassing the effects of the polymer parameters (within the range examined here) on the looping dynamics.

In contrast to the common view that dissociation rates are local phenomena only, these data suggest that the loop dissociation and association kinetics are both regulated by $J_{loop}$. We note that the looping J factor is sometimes interpreted as effective cohesive-end concentration or effective repressor concentration, and increasing effective concentration is thought to facilitate association kinetics (29, 30). However, this concept does not explain how the dissociation kinetics is modulated by an effective concentration: according to the simple kinetic framework discussed above, the looped-state lifetime should not depend on repressor concentration $[R]$. If we take the effective concentration interpretation of $J_{loop}$ literally, the fact that the dissociation kinetics, i.e., the looped lifetime shown in Fig. 3, depends on $J_{loop}$ is inconsistent with this framework. Explaining the dependence of the dissociation kinetics on $J_{loop}$ requires a different interpretation of $J_{loop}$ beyond effective concentration and more akin to how applied force distorts bonding free energy in force spectroscopy experiments (21–25). We first apply transition state theory to obtain some intuition about how $J_{loop}$ can modulate loop formation and breakdown rates, and then turn to a more sophisticated framework that more explicitly models the polymer mechanics. We note that our analysis based on free-energy landscapes is theoretically equivalent to expressing the effect of polymer deformation in terms of force and torque acting on the bond (5, 26, 27), and we used the free-energy treatment because of its conceptual simplicity.

The magenta curve in Fig. 2C shows a pathway between one unlooped state, where operator $A$ is bound, and the looped state with both operators bound. The transition state on this path has an unknown structure and a total free energy $F_{transition}$. The activation energies for the forward and reverse transitions are given by $\Delta F_{unloop}^\ddagger = (F_{transition} - F_{unlooped})$ and $\Delta F_{loop}^\ddagger = (F_{transition} - F_{looped}) = (\Delta F_{unloop}^\ddagger - \Delta F + E_B)$, where $\Delta F$ is the free energy of deforming the DNA into the looped state and $E_B$ is the favorable free energy of binding operator *B*. The loop formation and breakdown rates are given by transition state theory to be the following:

$$k_{on}^\beta = k_0^\beta e^{-\beta \Delta F_{unloop}^\ddagger} \quad [2]$$

and

$$k_{off}^\beta = k_0^\beta e^{-\beta \Delta F_{loop}^\ddagger} = k_0^\beta e^{-\beta\left(\Delta F_{unloop}^\ddagger - \Delta F + E_B\right)}, \quad [3]$$

where $k_0^\beta$ absorbs contributions from diffusivity and the shape of the free energy pathway. Similar equations can be derived for $k_{on}^\alpha$ and $k_{off}^\alpha$ (*SI Appendix*, section S2.3.2). Combining these with the equations given above and in *SI Appendix* for $\langle\tau_{unlooped}\rangle$ and $\langle\tau_{looped}\rangle$, we find that the mean unlooped and looped lifetimes will scale as follows:

$$\langle\tau_{unlooped}\rangle \propto e^{\beta \Delta F_{unloop}^\ddagger} \quad [4]$$

and

$$\langle\tau_{looped}\rangle \propto e^{\beta\left(\Delta F_{unloop}^\ddagger - \Delta F + E_B\right)}. \quad [5]$$

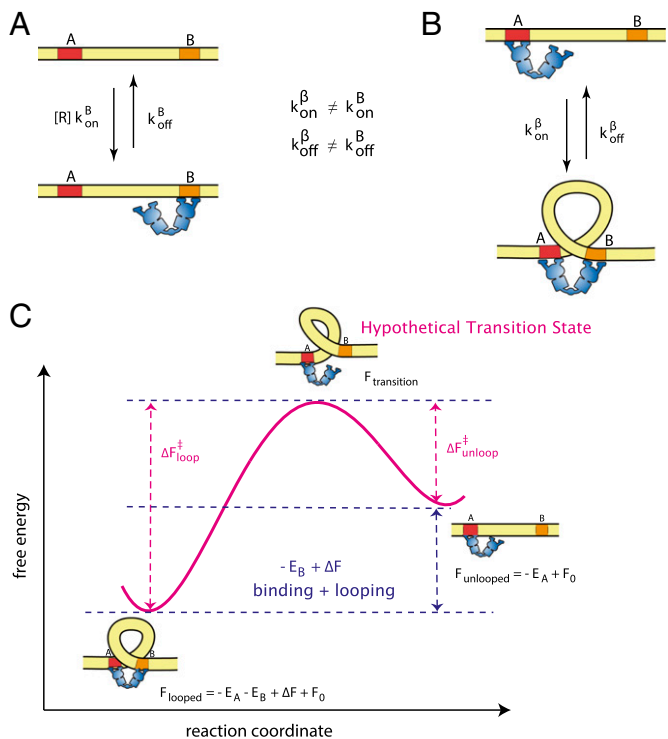

**Fig. 2.** Kinetic framework of protein-mediated looping. (*A*) Repressor association with operator *B* is a pseudo–first-order reaction with rate $[R]k_{on}^B$. Dissociation is a zeroth-order reaction with rate $k_{off}^B$. (*B*) Association of the repressor bound at operator *A* with the unbound operator *B* has a rate $k_{on}^\beta$, and dissociation from operator *B* has a rate $k_{off}^\beta$. Note that here we are distinguishing between the binding/unbinding of the first repressor head (rate constants $k_{on}^A, k_{off}^A, k_{on}^B, k_{off}^B$) and the binding/unbinding of the second repressor head to the same DNA (rate constants $k_{on}^\alpha, k_{off}^\alpha, k_{on}^\beta, k_{off}^\beta$). (*C*) The reaction curve (magenta) shows a hypothetical transition state, with unknown free energy, controls both the forward the reverse reactions. What is known from equilibrium measurements is the difference between the unlooped and looped states, $-E_B + \Delta F$ (binding energy plus the penalty for looping deformation).

Given the data in Fig. 3, we must conclude that not only $\Delta F_{unloop}^\ddagger$, but also $\Delta F_{loop}^\ddagger$, are determined by the DNA deformation energy, encapsulated in $\Delta F$. We note that the looped and unlooped state lifetimes in Fig. 3 *C* and *D* scale roughly linearly with *J* when plotted in log-log scale, and so we can obtain an approximate form of the relationship between $\tau$ and *J* (equivalently, $\Delta F$). We find that $\langle\tau_{unlooped}\rangle \propto J_{loop}^{-0.48 \pm 0.03}$, and $\langle\tau_{looped}\rangle \propto J_{loop}^{0.35 \pm 0.02}$. This provides some intuition into how far the unknown transition state is, in terms of the elastic deformation $\Delta F$, from the looped and unlooped states. Moreover, if the DNA chain were bent and twisted almost all of the way into the needed shape before binding of the second operator, the unlooped lifetime would scale as $J_{loop}^{-1}$ and the looped lifetime would be independent of $J_{loop}$ and chain length, a very different kinetic behavior from what we find here. This change in the $J_{loop}$ dependence could shift the kinetic lifetimes an order of magnitude over just one decade in loop length.

It is clear that this is not the case for our system; that is, our finding that both the looped and unlooped lifetimes are dependent upon $J_{loop}$ reveal that the process of going from the unlooped state to the transition state encompasses some, but not all, of the elastic deformation of the chain. Instead, the scaling exponents for $\tau$ versus *J* give an indication of the degree of release in elastic strain when moving from the looped state to the transition state. This model stands in contrast to previous models of ring closure that describe the DNA looping process as the first-passage time of the two ends coming within an approximately zero distance of separation (16–18).

To explore which elastic energies from the DNA and protein may be contributing to the reaction landscape of Fig. 2*C*, we have developed a molecular-level model for the looping and unlooping processes. We reason that the free-energy landscape has contributions from bending and twisting deformations of the DNA and a binding energy (between the free repressor binding domain and the empty operator) of finite interaction length, as described in Fig. 1*A*. This results in a distance from the looped state to the transition state, similar to the ideas of finite-scale chemical bonding in refs. 21 and 22.

The DNA loop region is modeled as a worm-like chain, which describes the polymer as an elastic thread that is subjected to thermal fluctuations (31). Although there is still considerable debate about the elastic nature of DNA at short lengths (32, 33), the worm-like chain model has a clear physical basis and its application resulted in reasonable value for the persistence length. The bending energy for a specific conformation is given by the following:

$$\beta E_{bend} = \frac{L_p}{2} \int_0^L ds \kappa(s)^2, \quad [6]$$

which depends on the square of the local curvature $\kappa(s)$. For a specific polymer conformation, the local curvature is equivalent to the inverse of the radius of a circle that is tangent to the curve (e.g., a straight chain segment has zero curvature and a tangent circle with infinite radius). This quadratic bending energy is consistent with linear elasticity theory of a thin elastic beam with a bending modulus $k_B T L_p$ (where $L_p$ is the persistence length). The polymer conformational free energy $F_{conf}(r)$ gives the free energy for fixing the end-to-end distance of the two operators to be *r*, incorporating both the bending deformation energy and the entropy of different DNA conformations. We show some example configurations from a Monte Carlo simulation in Fig. 4*A*. To determine $F_{conf}(r)$, we find the Green function $G(r)$, which gives the probability of the two ends being a distance *r* apart, by summing over all possible paths and weighting each by $e^{-\beta E_{bend}}$. We have previously derived the exact result for the Green function (34) and use this result to calculate the conformational free energy $\beta F_{conf}(r) = -\log[r^2 G(r)]$.

The binding of the operators to the Lac repressor requires proper orientational alignment between the binding face of the operators and the Lac binding domains. The intervening DNA length determines the undeformed orientation of the DNA helix at the operator, and proper alignment for binding incurs energetically costly twist deformation upon rotating the DNA into its proper orientation (which is in fact the origin of the modulation of *J* with loop length, noted in the text accompanying Fig. 3*A*). We define the twist angle $\theta$ to be the angle of rotation about the DNA axis at the unbound operator away from the ground-state untwisted angle (i.e., $\theta = 0$ is untwisted). We consider a simple model for the twist free energy $\beta F_{twist}(\theta) = (L_t/2L)\theta^2$, which is quadratic in the local twist deformation and evenly distributes the twist deformation over the length of the DNA between the operators. The twist persistence length $L_t$ represents the resistance to twist deformation. This model neglects the geometric coupling between twist and writhe of the chain, which becomes more relevant at longer chain lengths where out-of-plane conformations are not prohibited by the bending deformation energy (34, 35).

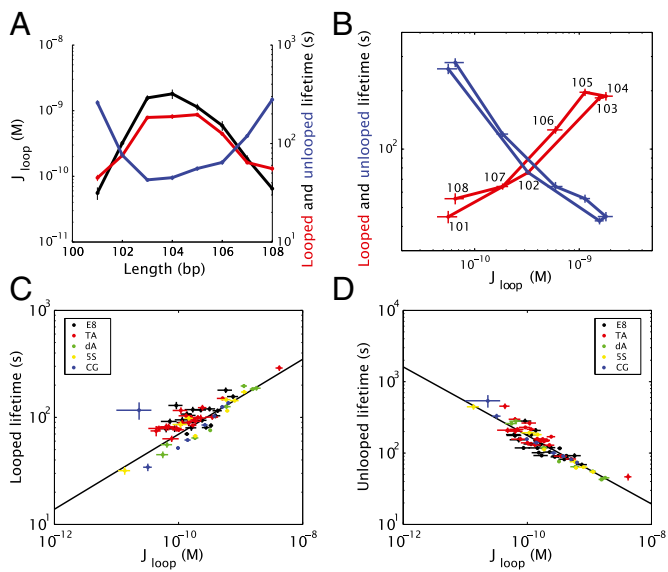

**Fig. 3.** Experimentally measured state lifetimes. (*A*) Looping *J* factor (black), mean unlooped lifetime (blue), and mean looped lifetime (red), for one helical period of the "dA" sequence. (*B*) The same data as in *A*, but with lifetimes plotted versus the *J* factor and the loop length (in base pairs) marked for the looped lifetime curve. (*C*) Mean looped state lifetime and (*D*) mean unlooped state lifetime versus $J_{loop}$, for one helical repeat each for three sequences ("dA," "5S," "CG"), and two helical repeats for two sequences ("E8," "TA").

The binding free energy, which drives the formation of the looped state, is modeled as a potential well with depth $\epsilon_0$ and an interaction length scale $\delta$. The separation of the two operators at the surface of the DNA strands is given by $r_a(r,\theta) = \sqrt{(r-a)^2 + a^2 - 2(r-a)a\cos(\theta - \theta_{op})}$, where *r* is the end-to-end distance of the DNA strands and $\theta$ is the twist angle at the unbound operator (see SI Appendix, Fig. S5, for an illustration). The preferred twist angle $\theta_{op} = 2\pi(L/L_{turn}) + \theta_0$ gives the twist angle that orients the empty operator to face the Lac repressor binding domain, where $\theta_0$ defines the twist angle necessary for docking DNA into Lac repressor even at integer DNA helical repeats. The representative images in Fig. 4*A* show how the DNA twists to orient the Lac repressor with the unbound operator as the ends are brought together. The DNA structure dictates the cross-sectional radius *a* (assumed to be $a = 1$ nm) (36) and the helical length $L_{turn}$ (assumed to be $L_{turn} = 10.46$ bp $\times 0.34$ nm) (34). The binding free energy $F_{bind}$ is then given by the following:

$$\beta F_{bind}(r,\theta) = \begin{cases} \dfrac{-2\epsilon_0}{1 + \exp\left[\dfrac{r_a(r,\theta)}{\delta}\right]}, & r > 2a, \\ \infty, & r \leq 2a, \end{cases} \quad [7]$$

which includes a steric cutoff at $r = 2a$ to account for the overlap of DNA backbone segments. This simple binding model aims to model the basic interaction between the DNA operator and Lac repressor by introducing only the binding affinity $\epsilon_0$ and the interaction length $\delta$ to capture the physical interaction. More detailed models of interaction could include more molecular detail, but our goal is to give the simplest representation of binding without introducing additional parameters that do not have well-defined values.

The three thermodynamic contributions $F_{conf}(r)$, $F_{twist}(\theta)$, and $F_{bind}(r,\theta)$ combine to give the total free-energy landscape $F_{total}(r,\theta)$, as shown in Fig. 4*A* for $L = 101$ bp and parameters $\epsilon_0 = 23.5$ (in $k_B T$ units), $\delta = 1.3$ nm, $L_p = 48$ nm, $L_t = 15$ nm, and $\theta_0 = 0.003\,\pi$. We find the minimum free-energy path from the looped state *X*, over the transition state *Y*, to the unlooped state *Z*, and plot each of the three free energy contributions, as well as the total free energy along this path, in Fig. 4*B*. From transition state theory, the looped lifetime is simply proportional to $e^{-\beta \Delta F^{\ddagger}_{loop}}$, and the unlooped lifetime is proportional to $e^{-\beta \Delta F^{\ddagger}_{unloop}}$, as given in Eqs. **4** and **5** above. We use the more sophisticated Fokker–Planck formalism and treat the reaction from the looped to unlooped state (and vice versa) as diffusion on a one-dimensional potential-energy landscape, given by $F_{total}(r)$ along the minimum free-energy path shown in Fig. 4*B*. Twist angle relaxation is much faster than changes in the end-to-end distance (37), and we avoid introducing an additional, poorly characterized parameter for the twist angle diffusivity by reducing the problem down to one dimension. We calculate the mean looped lifetime as the average first passage time from anywhere in $r < r_Y$ to leaving the transition state at $r = r_Y$, and similarly for the mean unlooped lifetime. The looping *J* factor is calculated from the polymer free energy difference:

$$J_{loop} = (1\text{ M})e^{-\beta \Delta F} = (1\text{ M})\exp\left[-\beta\left(F^{loop}_{poly} - F^{unloop}_{poly}\right)\right]. \quad [8]$$

Here, $F^{loop}_{poly}$ and $F^{unloop}_{poly}$ are calculated by averaging only the polymer elastic energies ($F_{conf} + F_{twist}$) (i.e., excluding the binding energy) over the end-to-end distance *r* smaller and larger than $r = r_Y$, respectively, with a Boltzmann weight given by $e^{-\beta F_{total}(r)}$. We refer the reader to SI Appendix, section S2.4, for more details.

To compare these theoretical results with the experimental results of Fig. 3, we find model parameters for the elastic parameters $L_p$, $L_t$, which could vary with DNA sequence, and the binding parameters $\epsilon_0$, $\delta$, and $\theta_0$, which should be consistent across all sequences with the same operators. We obtain $\theta_0 = 0.003\,\pi$ by looking at the peaks in the $J_{loop}$ data (see figure 2 in ref. 14), which occur when the twist-free orientation $2\pi L/L_{turn}$ is aligned with $\theta_0$. The model is able to reproduce the basic qualitative features of the data across a range of parameters, and we chose values of $\epsilon_0 = 23.5$ (in $k_B T$ units), and $\delta = 1.3$ nm as representative of a good fit to the data across all five sequences for the set of operators used here. These parameters are within the expected range, given the size of the Lac repressor arm (around 3–4 nm from the crystal structure) (8) and the binding energy of the repressor to DNA of approximately 16 $k_B T$ (38). We then varied the elastic parameters to find the best fit for each sequence, obtaining values of the persistence length $L_p$ ranging from 48 to 51 nm and the twist persistence length $L_t$ ranging from 10 to 70 nm. The values for $L_p$ are close to the canonical value for dsDNA of $L_p = 53$ nm (39). Although our twist persistence differs from the canonical value of $L_t = 110$ nm (37), we note that our twist model is much simplified. We do not include the details of the end orientations, twist angle entropic effects, and twist–writhe coupling, all of which could lead to the lower value of $L_t$ that we determined.

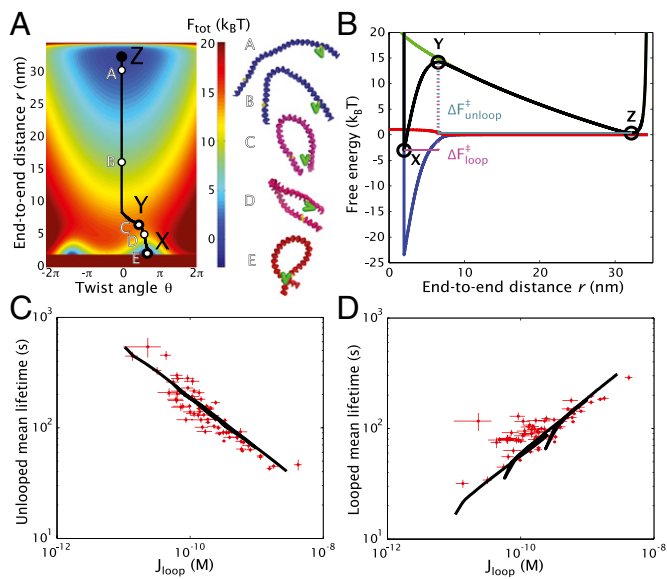

**Fig. 4.** Molecular model for DNA looping. (*A*) Total free-energy surface versus end-to-end distance $r$ and twist angle $\theta$. In this plot, $L = 101$ bp and the parameters are $\epsilon_0 = 23.5$ (in $k_BT$ units), $\delta = 1.3$ nm, $L_p = 48$ nm, $L_t = 15$ nm, and $\theta_0 = 0.003\,\pi$. The black curve indicates the minimum free-energy path between the looped state (*X*) and the unlooped state (*Z*), passing through the transition state (*Y*). Representative DNA conformations (as predicted by Monte Carlo simulation) at five different end-to-end separations are shown to the right of the free-energy surface, where the degree of twisting is indicated by the DNA coloration ranging from blue for $\theta = 0$ to red for $\theta = 0.678\,\pi$. (*B*) Free energy along the minimum free-energy path. The total free energy (black) is a combination of the polymer free energy (green), the twisting free energy (red), and the binding free energy (blue). The free-energy barriers to leave the looped and unlooped states are $\Delta F^{\ddagger}_{loop}$ and $\Delta F^{\ddagger}_{unloop}$, respectively. (*C*) Unlooped lifetime behavior. The experimentally determined unlooped lifetimes (red dots) are plotted versus $J_{loop}$, with the black line corresponding to the theoretical prediction as $L$ is varied from 89 to 115 bp. (*D*) Looped lifetime behavior. The looped lifetimes from the experiments (red dots) and theory (black line) are plotted using the same parameters as in *C*.

Theoretical predictions for the unlooped and looped mean lifetimes are shown in Fig. 4 *C* and *D*, using the same values as in Fig. 4*A*, and a full comparison with each sequence is given in *SI Appendix* (*SI Appendix*, Fig. S6). The theoretical lifetimes (black curves) exhibit an approximate power-law trend with $J_{loop}$ for the lengths ranging from 89 to 115 bp. Notably, the value of $J_{loop}$ exhibits both oscillations and an average increase as the length is increased from 89 to 115 bp. In this regard, the quantity $J_{loop}$ serves as a critical determinant of the looped and unlooped lifetimes.

One usually unrecognized feature introduced in our molecular model for looping is the treatment of the protein binding energy that has a well depth of $\epsilon_0$ and an interaction radius of $\delta$. These parameters are specifically dependent upon the properties of the protein and the operator binding interface. In addition to the data shown in Fig. 3, we also analyzed TPM trajectories with a different set of operators, specifically with the $O_1$ operator replaced by the slightly weaker $O_2$. These data are plotted as blue dots in Fig. 5. We have previously determined the energetic difference between these two operators to be 1.5 $k_BT$ (38). Because only one operator's affinity was changed, we would expect the resulting value of $\epsilon_0$ to be reduced by 0.75 $k_BT$. (We note that in these operator-swap experiments, the sequence of the loop was somewhat altered as well, but not its total length, and as such we expect most or all of the change to be due to the difference in the binding well depth $\epsilon_0$). Our model prediction, given in black in Fig. 5, clearly agrees well with the experimental results when $\epsilon_0$ is reduced by 0.75 $k_BT$ and all other parameters are kept the same as in Fig. 4.

Consistent with our theory, only the looped lifetimes are affected by the change of operator. In the free-energy plot in Fig. 4*B*, we see that the binding energy (blue) only begins to affect the total free-energy curve (black) once the configuration is to the left of the transition state; i.e., it is in the looped state. Likewise, we see that the twist energy (red) only begins at end-to-end distances less than the transition state end-to-end distance $r_Y$. Thus, our molecular-level model has given us clear insight into the elastic deformation present at the transition state, and this agrees well with our experimental measurements.

The other major parameter introduced for the binding energy was a finite length scale for the DNA–protein interaction. This parameter is critical to explain our findings that both the looped and unlooped lifetimes depend upon the $J$ factor. The finite length scale of interaction $\delta$ affects the transition state $Y$ by changing the end-to-end distance and twist angle at which this state occurs. For large $\delta$, the transition state would occur at a farther end-to-end distance and thus exhibit a notable release of deformation energy compared with the looped state $X$, leading to a dependence of the looped lifetime on $J_{loop}$. This parameter $\delta$ phenomenologically models both the size (8) and flexibility (40) of the protein mediating the loop, and could also account for other effects such as electrostatic interactions or nonspecific binding leading to sliding along the DNA chain (41). Thus, in experiments with proteins of smaller size or flexibility than Lac repressor, we would expect a decreased scaling exponent and hence decreased dependence of the looped lifetime on $J_{loop}$. We will explore the effect of this interaction distance further in an upcoming manuscript.

## Conclusion

Using the single-molecule technique of tethered particle motion to examine looping and unlooping lifetimes by the classic Lac repressor looping protein, we have shown here that both the looped and unlooped lifetimes depend upon the $J$ factor, indicating that the dissociation rate is dependent upon the DNA and protein elasticity. These findings are unexpected based on the common treatment of the $J$ factor as an effective protein concentration, and have been ignored by previous studies of DNA looping. We also note that the $J$ factor-modulated state lifetimes, having a 1- to 10-min dynamical range, are comparable to *E. coli*'s cell division time. The state lifetimes are sensitive to how they scale with $J$, and within a decade of loop length variation the response times can change an order of magnitude. It is therefore interesting to explore how DNA mechanics modulates the in vivo looping and unlooping rates and assess its influence on how individual cells respond to nutrient fluctuations. To explain our experimental results, we have developed a molecular-level model that accounts for the role of both the polymer and protein deformation in DNA looping and unlooping kinetics. This model includes a simple but straightforward and physically derived picture for the three energies necessary to explain looping in short, stiff chains: bending, twisting, and binding. The binding energy used allows us to incorporate the protein elasticity through the introduction of a finite length scale of interaction that modulates the degree of favorable binding depending upon the end-to-end distance. We find the lifetimes calculated from this model to be in good agreement with our experimental results with realistic physical parameters, and that the model provides additional insights into the properties of the transition state and how the elastic energy changes during the course of the looping reaction. Finally, we note that long-range ordering of opening and closing kinetics by the system's free energy landscape should be a general framework that goes beyond the DNA or repressor-specific variables, and can be applied to other elastic systems such as ligand–receptor reaction (42) or protein assembly (43), where flexible tethers are important for the biological functions.

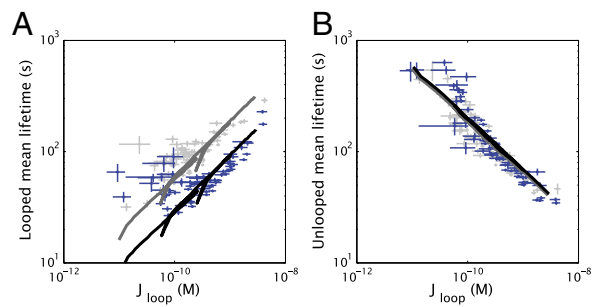

**Fig. 5.** Changing operator affinity shifts looped lifetimes. (*A*) Looped lifetime behavior. The experimentally determined looped lifetimes (blue) for all five sequences with $O_{id}$ and $O_2$ operators, instead of $O_{id}$ and $O_1$ as in Fig. 3, are plotted versus $J_{loop}$, with the black line corresponding to the theoretical prediction as $L$ is varied from 89 to 115 bp. All of the model parameters are the same as in Fig. 4 except $\epsilon_0$ has been reduced to 22.75 $k_BT$. For reference, the light gray dots are the data from the $O_{id}$ and $O_1$ operators, and the dark gray line is the theory curve from Fig. 4. (*B*) Unlooped lifetime behavior. The unlooped lifetimes for the $O_{id}$ and $O_2$ operators (blue) and theory (black line) are plotted, the same as in *A*.

## Materials and Methods

TPM experiments were performed as previously described (12, 14). Briefly, a micrometer-sized bead is tethered to one end of a linear DNA, with the other end attached to a microscope coverslip. The motion of the bead depends on the effective length of the DNA such that loop formation induced by the binding of the Lac repressor to its two operators results in a quantifiable change of the bead's motion. We record the trajectory of looping and unlooping for each DNA molecule, under various experimental conditions such as Lac repressor concentration and DNA sequence. We use a thresholding procedure to quantify the looped and unlooped lifetimes from the recorded trajectories. Details of our implementation of the half-amplitude thresholding procedure are given in *SI Appendix*, section S1, and a comparison of our results to those in previous studies using TPM to measure Lac repressor looping and unlooping rates, showing good agreement between our results and these previous studies, is given in *SI Appendix*, section S3.2. Experimental errors are reported as SEs on the means, calculated according to the bootstrapping method described in *SI Appendix*, section S1.

**ACKNOWLEDGMENTS.** We are grateful to Martin Lindén, Justin Bois, Mattias Rydenfelt, Yun Mou, James Boedicker, Robert Brewster, and Arbel Tadmor for helpful discussions and comments, and to David Wu, David Van Valen, Heun Jin Lee, Geoff Lovely, Hernan Garcia, Franz Weinert, Chao Liu, Luke Breuer, and Matthew Johnson for help with experimental setup and analysis. This work was supported by the National Institutes of Health [Grants DP1 OD000217A (Directors Pioneer Award), R01 GM085286, R01 GM085286-01S1, and 1 U54 CA143869 (Northwestern Physical Sciences–Oncology Center)], the National Science Foundation through a graduate fellowship (to S.J.) and through a grant from the National Science Foundation (PHY-1305516) (to P.M. and A.J.S.), and the Fondation Pierre Gilles de Gennes (R.P.).

1. Griffith J, Hochschild A, Ptashne M (1986) DNA loops induced by cooperative binding of lambda repressor. *Nature* 322(6081):750–752.
2. Zeller RW, et al. (1995) A multimerizing transcription factor of sea urchin embryos capable of looping DNA. *Proc Natl Acad Sci USA* 92(7):2989–2993.
3. Peters JP, 3rd, Maher LJ (2010) DNA curvature and flexibility in vitro and in vivo. *Q Rev Biophys* 43(1):23–63.
4. Li G-W, Berg O, Elf J (2009) Effects of macromolecular crowding and DNA looping on gene regulation kinetics. *Nat Phys* 5:294–297.
5. Laurens N, et al. (2012) DNA looping by FokI: The impact of twisting and bending rigidity on protein-induced looping dynamics. *Nucleic Acids Res* 40(11):4988–4997.
6. Gross P, et al. (2011) Quantifying how DNA stretches, melts and changes twist under tension. *Nat Phys* 7:731–736.
7. Vafabakhsh R, Ha T (2012) Extreme bendability of DNA less than 100 base pairs long revealed by single-molecule cyclization. *Science* 337(6098):1097–1101.
8. Lewis M, et al. (1996) Crystal structure of the lactose operon repressor and its complexes with DNA and inducer. *Science* 271(5253):1247–1254.
9. Schafer DA, Gelles J, Sheetz MP, Landick R (1991) Transcription by single molecules of RNA polymerase observed by light microscopy. *Nature* 352(6334):444–448.
10. Finzi L, Gelles J (1995) Measurement of lactose repressor-mediated loop formation and breakdown in single DNA molecules. *Science* 267(5196):378–380.
11. Nelson PC, et al. (2006) Tethered particle motion as a diagnostic of DNA tether length. *J Phys Chem B* 110(34):17260–17267.
12. Johnson S, Lindén M, Phillips R (2012) Sequence dependence of transcription factor-mediated DNA looping. *Nucleic Acids Res* 40(16):7728–7738.
13. Crothers DM, Drak J, Kahn JD, Levene SD (1992) DNA bending, flexibility, and helical repeat by cyclization kinetics. *Methods Enzymol* 212:3–29.
14. Johnson S, Chen Y-J, Phillips R (2013) Poly(dA:dT)-rich DNAs are highly flexible in the context of DNA looping. *PLoS One* 8(10):e75799.
15. Vanzi F, Broggio C, Sacconi L, Pavone FS (2006) Lac repressor hinge flexibility and DNA looping: Single molecule kinetics by tethered particle motion. *Nucleic Acids Res* 34(12):3409–3420.
16. Jun S, Bechhoefer J, Ha B-Y (2003) Diffusion-limited loop formation of semiflexible polymers: Kramers theory and the intertwined time scales of chain relaxation and closing. *Europhys Lett* 64:420–426.
17. Hyeon C, Thirumalai D (2006) Kinetics of interior loop formation in semiflexible chains. *J Chem Phys* 124(10):104905.
18. Toan NM, Morrison G, Hyeon C, Thirumalai D (2008) Kinetics of loop formation in polymer chains. *J Phys Chem B* 112(19):6094–6106.
19. Van Valen D, Haataja M, Phillips R (2009) Biochemistry on a leash: The roles of tether length and geometry in signal integration proteins. *Biophys J* 96(4):1275–1292.
20. Reeves D, Cheveralls K, Kondev J (2011) Regulation of biochemical reaction rates by flexible tethers. *Phys Rev E Stat Nonlin Soft Matter Phys* 84(2 Pt 1):021914.
21. Bell GI (1978) Models for the specific adhesion of cells to cells. *Science* 200(4342):618–627.
22. Merkel R, Nassoy P, Leung A, Ritchie K, Evans E (1999) Energy landscapes of receptor-ligand bonds explored with dynamic force spectroscopy. *Nature* 397(6714):50–53.
23. Dudko OK, Hummer G, Szabo A (2008) Theory, analysis, and interpretation of single-molecule force spectroscopy experiments. *Proc Natl Acad Sci USA* 105(41):15755–15760.
24. Liphardt J, Onoa B, Smith SB, Tinoco I, Jr, Bustamante C (2001) Reversible unfolding of single RNA molecules by mechanical force. *Science* 292(5517):733–737.
25. Mihardja S, Spakowitz AJ, Zhang Y, Bustamante C (2006) Effect of force on mono-nucleosomal dynamics. *Proc Natl Acad Sci USA* 103(43):15871–15876.
26. Joseph C, Tseng C-Y, Zocchi G, Tlusty T (2014) Asymmetric effect of mechanical stress on the forward and reverse reaction catalyzed by an enzyme. *PLoS One* 9(7):e101442.
27. Le TT, Kim HD (2014) Probing the elastic limit of DNA bending. *Nucleic Acids Res* 42(16):10786–10794.
28. Chen YF, Milstein JN, Meiners JC (2010) Femtonewton entropic forces can control the formation of protein-mediated DNA loops. *Phys Rev Lett* 104(4):048301.
29. Law SM, Bellomy GR, Schlax PJ, Record MT, Jr (1993) In vivo thermodynamic analysis of repression with and without looping in lac constructs. Estimates of free and local lac repressor concentrations and of physical properties of a region of supercoiled plasmid DNA in vivo. *J Mol Biol* 230(1):161–173.
30. Krishnamurthy VM, Semetey V, Bracher PJ, Shen N, Whitesides GM (2007) Dependence of effective molarity on linker length for an intramolecular protein-ligand system. *J Am Chem Soc* 129(5):1312–1320.
31. Saitô N, Takahashi K, Yunoki Y (1967) The statistical mechanical theory of stiff chains. *J Phys Soc Jpn* 22:219–226.
32. Wiggins PA, et al. (2006) High flexibility of DNA on short length scales probed by atomic force microscopy. *Nat Nanotechnol* 1(2):137–141.
33. Geggier S, Vologodskii A (2010) Sequence dependence of DNA bending rigidity. *Proc Natl Acad Sci USA* 107(35):15421–15426.
34. Spakowitz AJ (2006) Wormlike chain statistics with twist and fixed ends. *Europhys Lett* 73:684–690.
35. Shimada J, Yamakawa H (1984) Ring-closure probabilities for twisted wormlike chains. Application to DNA. *Macromolecules* 17:689–698.
36. Sinden RR (1994) *DNA Structure and Function* (Academic, San Diego).
37. Bryant Z, et al. (2003) Structural transitions and elasticity from torque measurements on DNA. *Nature* 424(6946):338–341.
38. Garcia HG, Phillips R (2011) Quantitative dissection of the simple repression input-output function. *Proc Natl Acad Sci USA* 108(29):12173–12178.
39. Smith SB, Finzi L, Bustamante C (1992) Direct mechanical measurements of the elasticity of single DNA molecules by using magnetic beads. *Science* 258(5085):1122–1126.
40. Czapla L, Grosner MA, Swigon D, Olson WK (2013) Interplay of protein and DNA structure revealed in simulations of the lac operon. *PLoS One* 8(2):e56548.
41. Kalodimos CG, et al. (2004) Structure and flexibility adaptation in nonspecific and specific protein-DNA complexes. *Science* 305(5682):386–389.
42. Jeppesen C, et al. (2001) Impact of polymer tether length on multiple ligand-receptor bond formation. *Science* 293(5529):465–468.
43. Zappulla DC, Cech TR (2004) Yeast telomerase RNA: A flexible scaffold for protein subunits. *Proc Natl Acad Sci USA* 101(27):10024–10029.